\documentclass[journal]{IEEEtran}

\usepackage{graphicx}
\usepackage{amsmath,amssymb,bm}
\usepackage{booktabs}
\usepackage{csquotes}
\usepackage{yhmath}
\usepackage{siunitx}
\usepackage{enumitem}
\usepackage{etoolbox}
\usepackage{url}
\usepackage[shortcuts,acronym]{glossaries}
\usepackage{nomencl} \makenomenclature
\usepackage{cite}
\usepackage[hidelinks]{hyperref} % last

\AtBeginDocument{\DeclareSIUnit{\MWh}{MWh}}
\AtBeginDocument{\DeclareSIUnit{\EUR}{EUR}}
\AtBeginDocument{\DeclareSIUnit{\a}{a}}
\AtBeginDocument{\DeclareSIUnit{\MW}{MW}}
\newcommand {\R}    {\mathbb{R}}

\newcommand	{\charge}	{\mathrm{char}}
\newcommand	{\dis}	{\mathrm{dis}}
\newcommand	{\zyk}	{\mathrm{cyc}}

\newcommand{\daa}   {\mathrm{DAA}}
\newcommand{\ida}   {\mathrm{IDA}}
\newcommand{\idc}   {\mathrm{IDC}}
\newcommand{\init}   {\mathrm{init}}

\newglossarystyle{listspaced}{%
  \setglossarystyle{list}
    {\begin{description}[leftmargin=1em, font=\normalfont]}{\end{description}}}

\makeglossaries % generates the acronym list

\newacronym{mpc}{MPC}{Model Predictive Control}
\newacronym{empc}{EMPC}{Economic Model Predictive Control}
\newacronym{soc}{SOC}{State of Charge}
\newacronym{bess}{BESS}{Battery Energy Storage System}
\newacronym{milp}{MILP}{Mixed Integer Linear Programming}
\newacronym{rh}{RH}{Rolling Horizon}
\newacronym{sdp}{SDP}{Stochastic Dynamic Programming}
\newacronym{sddp}{SDDP}{Stochastic Dual Dynamic Programming}
\newacronym{dp}{DP}{Dynamic Programming}
\newacronym{nn}{NN}{Neural Network}
\newacronym{lstm}{LSTM}{Long Short-Term Memory}
\newacronym{arma}{ARMA}{Auto Regressive-Moving Average}
\newacronym{porfc}{PORFC}{Parametric Online RainFlow-Counting}
\newacronym{pdf}{PDF}{Probability Density Forecasting}
\newacronym{lfp}{LFP}{Lithium Iron Phosphate}
\newacronym{nmc}{NMC}{Lithium Nickel Manganese Cobalt}
\newacronym{tms}{TMS}{Thermal Management System}
\newacronym{bms}{BMS}{Battery Management System}
\newacronym{h2}{H}{Hydrogen}
\newacronym{npv}{NPV}{Net Present Value}
\newacronym{smpc}{SMPC}{Stochastic Model Predictive Control}
\newacronym{lp}{LP}{Linear Programming}
\newacronym{daa}{DAA}{Day Ahead Auction}
\newacronym{ida}{IDA}{Intraday Auction}
\newacronym{idc}{IDC}{Intraday Continuous}
\newacronym{eex}{EEX}{European Energy Exchange}
\newacronym{tso}{TSO}{Transmission System Operator}
\newacronym{mae}{MAE}{Mean Absolute Error}
\newacronym{rmse}{RMSE}{Root Mean Square Error}
\newacronym{epex}{EPEX}{European Power Exchange}

\renewcommand\nomgroup[1]{%
  \ifstrequal{#1}{A}{\item[Parameters]}{%
  \ifstrequal{#1}{C}{\vspace{6pt}\item[Other]}{%
  \ifstrequal{#1}{D}{\vspace{6pt}\item[Other]}{%
  \ifstrequal{#1}{B}{\vspace{6pt}\item[Variables]}{}}}}%
}

\begin{document}
\title{Optimized Operation of Standalone Battery Energy Storage Systems in the Cross-Market \\ Energy Arbitrage Business}

\author{Luis~van~Sandbergen
\thanks{L. van Sandbergen is with the Institute of Energy Systems and Energy Management, Ruhr West University of Applied Sciences (HRW), Bottrop, NRW, Germany (e-mail: luis.vansandbergen@stud.hs-ruhrwest.de).}
\thanks{An earlier version of this paper was submitted in August 2024 as part of the “Discovery-led Project” course at HRW, carried out in cooperation with Iqony GmbH. This version contains minor revisions prior to arXiv publication.}
}

\maketitle

\begin{abstract}
The provision of renewable electricity is the foundation for a sustainable future. 
To achieve the goal of sustainable renewable energy, Battery Energy Storage Systems (BESS) could play a key role to counteract the intermittency of solar and wind generation power.
In order to aid the system, the BESS can simply charge at low wholesale prices and discharge during high prices, which is also called energy arbitrage.
However, the real-time execution of energy arbitrage is not straightforward for many companies due to the fundamentally different behavior of storages compared to conventional power plants.
In this work, the optimized operation of standalone BESS in the cross-market energy arbitrage business is addressed by describing a generic framework for trading integrated BESS operation, the development of a suitable backtest engine and a specific optimization-based strategy formulation for cross-market optimized BESS operation.
In addition, this strategy is tested in a case study with a sensitivity analysis to investigate the influence of forecast uncertainty.
The results show that the proposed strategy allows an increment in revenues by taking advantage of the increasing market volatility. Furthermore, the sensitivity analysis shows the robustness of the proposed strategy, as only a moderate portion of revenues will be lost if real forecasts are adopted.
\end{abstract}


\IEEEpeerreviewmaketitle

\printglossary[type=\acronymtype,title={List of Abbreviations},  style=listspaced, nonumberlist, nogroupskip] 

\printnomenclature

\nomenclature[A]{$\Delta t$}{Time resolution (h)}
\nomenclature[A]{$Q$}{Number of all quarters in test period}
\nomenclature[A]{$Q^{+}$}{Number of all quarters in test period plus 1}
\nomenclature[A]{$H$}{Number of all hours in test period}
\nomenclature[A]{$H^{+}$}{Number of all hours in test period plus 1}
\nomenclature[A]{$N_{p}$}{Prediction horizon}
\nomenclature[A]{$P^{\max}$}{Maximum charge/discharge power (MW)}
\nomenclature[A]{$E^{\max}$}{Maximum amount of energy (MWh)}
\nomenclature[A]{$\gamma$}{Coefficient for self-discharge rate}
\nomenclature[A]{$\eta^{\charge}$}{Charging efficiency}
\nomenclature[A]{$\eta^{\dis}$}{Discharging efficiency}
%\nomenclature[C]{$\bm{x}$}{Vector of decision variables}
\nomenclature[A]{$\widehat{c^{\daa}}$}{Price forecast for DAA (€/MWh)}
\nomenclature[B]{$P^{\daa}_{\charge}$}{Charging power on DAA (MW)}
\nomenclature[B]{$P^{\daa}_{\dis}$}{Discharging power on DAA (MW)}
\nomenclature[A]{$E_{\init}$}{Initial State of Charge (MWh)}
\nomenclature[B]{$E$}{Storages energy amount (MWh)}
\nomenclature[A]{$n^{\zyk}$}{Daily allowed full equivalent cycles}
\nomenclature[A]{$\widehat{c^{\ida}}$}{Price forecast for IDA (€/MWh)}
\nomenclature[B]{$P^{\ida}_{\charge}$}{Charging power on IDA (MW)}
\nomenclature[B]{$P^{\ida}_{\dis}$}{Discharging power on IDA (MW)}
\nomenclature[B]{$\overline{P^{\ida}_{\charge}}$}{Charging power on IDA to close position (MW)}
\nomenclature[B]{$\overline{P^{\ida}_{\dis}}$}{Discharging power on IDA to close position (MW)}
\nomenclature[A]{$\widehat{c^{\idc}}$}{Price forecast for IDC (€/MWh)}
\nomenclature[B]{$P^{\idc}_{\charge}$}{Charging power on IDC (MW)}
\nomenclature[B]{$P^{\idc}_{\dis}$}{Discharging power on IDC (MW)}
\nomenclature[B]{$\overline{P^{\idc}_{\charge}}$}{Charging power on IDC to close position (MW)}
\nomenclature[B]{$\overline{P^{\idc}_{\dis}}$}{Discharging power on IDC to close position (MW)}
\nomenclature[A]{$P^{\daa,\ida,\idc}_{\charge}$}{Existing charge (buys) position from DAA, IDA and IDC}
\nomenclature[A]{$P^{\daa,\ida,\idc}_{\dis}$}{Existing discharge (sells) position from DAA, IDA and IDC}
\nomenclature[B]{$X$}{Binary variable: 1 for charging, 0 for discharging}
\nomenclature[B]{$h$}{Time step for hours}
\nomenclature[B]{$q$}{Time step for quarter-hours}

\newpage
\section{Introduction}
\IEEEPARstart{T}{he} provision of renewable electricity is the foundation for a sustainable future. Particularly, when coupled with electric vehicles, heat-pump systems, or any other low-carbon electrical technology, this can result in a substantial decrease in greenhouse gas emissions. Unfortunately, current renewable generation technology, such as solar and wind, experience huge intermittency in generating power. This leads to a mismatch between demand and supply. To overcome this challenge, two possible paths are available as of current knowledge. The first is to let the demand follow the generation, which is also known as Demand-Side-Management. 
Nonetheless, it is not possible for all consumers to reduce their electricity consumption, especially when considering the constant operation of hospitals or industrial processes. The second path is to store the generated energy and provide it when needed.

A promising storage technology are Battery Energy Storage Systems (\acs{bess}). These storages are often based on lithium-ion-cells and can reach high power levels as well as storage capacities up to multiple gigawatt-hours. A lithium-ion cell consists mainly of two electrodes, called anode and cathode, and an ionically conductive electrolyte, with lithium ions migrating between the electrodes during charging and discharging. Compared to other batteries, they are characterized by a high cycle stability and a low self-discharge rate. With the right operation, this technology is both safe and comparatively friendly for the environment. \cite{Kurzweil:2018}

The company Iqony GmbH has over a decade of experience operating BESS. 
This started with the pilot project “Lithium-Ion-Electricity-Storage-System” (LESSY) in 2009 and was extended by another \SI{90}{\MW} and \SI{120}{\MWh} storage system in 2016 \cite{Iqony:2023a}. 
Currently, the company has another BESS project under development with a capacity of \SI{50}{\MW} and \SI{250}{\MWh} \cite{Iqony:2024a}.
Their experience extends to engineering multiple battery storage projects and conducting feasibility studies for industrial customers. In addition, the company has already produced several scientific works on the subject \cite{Lehmann:2022, Lehmann:2018, Hidalgo:2020, vanSandbergen:2024b}.

Iqony's BESS have so far been used to provide primary control power.
Therefore, the company focused its efforts on this particular application.
However, a more widespread application, such as energy arbitrage, may now be economically viable. 
With the utilization of energy storage as a flexible option, energy arbitrage is the purchase of energy during periods of low electricity prices and the sale of energy during periods of high prices.
Energy arbitrage is gaining attention due to the demand limit in the balancing market and the rising volatility of the electricity prices on the wholesale market in recent years.
Still, the issue of optimizing battery operation in energy arbitrage poses a challenge for many companies, due to the fundamentally different characteristics in the operation of storages compared to conventional power plants.

Several approaches for the optimal operation of BESS from the literature have been investigated \cite{Lehmann:2022, Collath:2023, Kumtepeli:2020, Metz:2018, Hashmi:2019, Xie:2021, Krishnamurthy:2018, Abdulla:2018, Hafiz:2020, Pelzer:2016, Abramova:2021,Loew:2021}, with Mixed-Integer Linear Programming (\acs{milp}) in a Rolling Horizon (\acs{rh}) framework found to be particularly effective.
\begin{table*}[ht]
\renewcommand{\arraystretch}{1.2}
\caption{Literature Overview on Optimal BESS Operation}
\label{table:literature}
\centering
\begin{tabular}{*{7}{c}}
\hline
Reference & Method & Scale\footnotemark{} & Storage Type\footnotemark{} & Application & Price prediction & Configuration \\
\hline
Lehmann et al. \cite{Lehmann:2022} & MILP & Utility & Li-Ion/NMC & Primary control & NN & Standalone\\
Collath et al. \cite{Collath:2023} & MPC/MILP & Utility & Li-Ion/LFP & Arbitrage & - & Standalone\\
Kumtepeli et al. \cite{Kumtepeli:2020} & MPC/MILP & Industrial &  Li-Ion/LPF & Arbitrage & - & Standalone\\
Metz et al. \cite{Metz:2018} & MIP & Utility & Li-Ion & Arbitrage & - & Standalone \\
Hashmi et al. \cite{Hashmi:2019} & MPC/\acs{lp} & Residential & Battery & Arbitrage & ARMA & BESS/PV/Load \\
Xie et al. \cite{Xie:2021} & MPC/MILP & Utility & Battery & (Arbitrage) & (Uncertainty sets) & BESS/Wind\\
Loew et al. \cite{Loew:2021} & MPC/PORFC & Residential & Li-Ion & Primary control & - & Standalone\\
Hafiz et al. \cite{Hafiz:2020} & RH/\acs{sddp} & Residential & Battery & Cost reduction & NN/LSTM & BESS/PV/Load\\
Abdulla et al. \cite{Abdulla:2018} & RH/SDP & Residential & Li-Ion & Cost reduction & - & BESS/PV/Load\\
Krishnamurthy et al. \cite{Krishnamurthy:2018} & MILP & Utility & Battery & Arbitrage & ARIMA & Standalone\\
\hline
\end{tabular}%}
\end{table*}
The Iqony related work of Lehmann et al. \cite{Lehmann:2018} explored the optimized operation of large-scale battery systems using mathematical optimization and neural networks for primary control. 
This work shows the potential of advanced computational methods in enhancing the efficiency of BESS operations. The authors identified three key research areas: modeling aging processes, multi-use optimization frameworks, and managing uncertainties in real-time.
Collath et al. \cite{Collath:2023} aimed to increase the lifetime profitability of BESS through aging-aware operation. They developed a Model Predictive Control (\acs{mpc}) framework incorporating linearized MILP models for lithium-iron-phosphate (\acs{lfp}) cell degradation. Their study demonstrated a significant increase in lifetime profit by considering aging costs based on system costs and electricity price profiles. 
This work underscores the importance of accurate aging models in optimizing BESS operations.
Kumtepeli et al. \cite{Kumtepeli:2020} presented a 3D-MILP model for optimizing energy arbitrage with battery storage, considering electro-thermal performance and semi-empirical aging models. Their approach, implemented in an MPC framework, highlighted the benefits of detailed aging models and the integration of thermal dynamics. The study revealed that temperature and SOC awareness significantly enhance the BESS's lifespan and profitability.
Metz and Saravia \cite{Metz:2018} investigated the use of BESS for price arbitrage in the German 15- and 60-minute intraday markets. Their mixed-integer optimization framework accounted for negative electricity prices and dual lifetime limitations.
Hashmi et al. \cite{Hashmi:2019} addressed the optimal storage arbitrage problem using linear programming (\acs{lp}) within an MPC framework, incorporating Auto Regressive-Moving Average (\acs{arma}) models for predicting electricity prices, consumer loads, and renewable generation. While their framework was tailored for residential applications, it emphasized the importance of considering uncertainties in electricity price forecasts.
Xie et al. \cite{Xie:2021} developed a robust MPC-based bidding strategy for wind storage systems in real-time energy and regulation markets. Their framework accounted for forecast uncertainties in wind energy and market prices, demonstrating superior performance compared to stochastic optimization methods. The study shows the potential of robust MPC in managing uncertainties and optimizing multi-use strategies.

Several other studies have explored different aspects of BESS optimization.
Loew et al. \cite{Loew:2021} used economic MPC to manage lithium-ion battery aging via online Rain Flow-analysis (\acs{porfc}), while Hafiz et al. \cite{Hafiz:2020} employed real-time stochastic optimization with deep learning-based forecasts for residential PV applications. Abdulla et al. \cite{Abdulla:2018} utilized Stochastic Dynamic Programming (\acs{sdp}) to account for battery degradation in energy storage systems. Krishnamurthy et al. \cite{Krishnamurthy:2018} proposes a stochastic formulation for energy storage arbitrage under day ahead and real-time price uncertainty.
Table \ref{table:literature} shows an overview of publications that propose operation strategies with BESS.
These publications collectively highlight the diverse methodologies and applications of BESS optimization in the literature.

Despite the extensive research, to the best of the authors' knowledge a comprehensive trading-integrated framework for real-time utility scale BESS operation is still missing. 
Accordingly, the aim of this work is to develop a trading integrated framework and strategy for optimally operating a standalone BESS for energy arbitrage. 
The paper's primary contributions are:
1) the description of a general trading integrated BESS operation framework,
2) the development of a backtest engine for specific operation strategies,
3) the formulation of an optimization-based strategy for BESS bidding,
4) a case study with evaluation of a \SI{10}{\MW}/\SI{10}{{\MWh}} storage system and investigation of forecast uncertainty.

The remainder of the paper at hand is structured as follows: 
First, in Section \ref{operation_framework} the general framework for BESS operation is described. A developed backtest engine based on this framework is presented in Section \ref{backtest_engine}.
An example mathematical strategy for BESS operation is afterwards formulated in Section \ref{stategy_formulation}.
In Section \ref{case_study} the backtest engine and developed strategy is then tested in a case study. Finally, Section \ref{Conclusion} concludes the paper and provides an outlook for future research.

\footnotetext{The scale corresponds to roughly residential $\approx$ 10 kWh, industrial $\approx$ 100 kWh and utility $>$1 MWh.}
\footnotetext{The battery category applies to cases without specific information.}

%
%------------------------------------------------------------------------------------
%
\section{Framework for Energy Trading Integrated BESS Operation} \label{operation_framework}

\begin{figure}[htp]
    \centering 
    \includegraphics{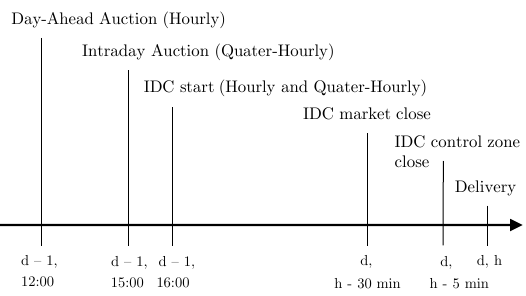}
    \caption{A timeline of the daily routine of the German spot electricity market. d, h represent the day and hour of delivery (based on \cite{NARAJEWSKI:2022}).}
    \label{fig:daily_routine}
\end{figure}

On the European Energy Exchange (\acs{eex}) there exist mainly three markets for delivery in the German Transmission System Operator (\acs{tso}) zones: the futures, spot, and balancing markets. 
The biggest and most important one is the European Power Exchange (\acs{epex}) Spot market, which is used for the physical trading of electricity, while the futures market's purpose is for financial hedging transactions \cite{NARAJEWSKI:2022}.
The EPEX spot market mainly consists of the Day Ahead Auction (\acs{daa}) and Intraday Auction (\acs{ida}), as well as the Intraday Continuous (\acs{idc}) trading.
The Day Ahead Auction is hourly and closes at 12:00 on the day before delivery. 
The order book opens 45 days in advance, and the electricity is traded in 24-hour intervals.
The Intraday Auction is quarter-hourly and, despite its name, closes at 15:00 on the previous day.\footnotemark{}  
The electricity is traded in 96 quarter-hour intervals, and the order books also open 45 days in advance.
Both auctions follow the “pay-as-cleared” principle, meaning an hourly or quarter-hourly clearing price is calculated by the exchange from the placed bids (see \cite{NEMO:2024}), which is then paid to all participants whose bids are accepted after the submission deadline.
The Intraday Continuous market fully starts at 16:00 on the day preceding the delivery, and the products are tradable until 1 hour before delivery Europe-wide (XBID), 30 min before delivery in Germany, or up to 5 min before delivery in the respective TSO control zone (Amprion, TenneT, 50Hertz and TransnetBW).
Between two neighboring countries, the lead time is one hour.
The electricity can be traded in 24 hours, 48 half-hours or 96 quarter-hours intervals.
The intraday continuous market has recently gained a steady increase and thus gained importance compared to the DAA \cite{Baule:2021}.
Unbalanced positions have to be subsequently balanced on the balancing market at typically high prices. 
Although the balancing market is interesting for BESS trading due to its high volatility, active trading on the market is prohibited in Germany.
There are minimum volume increments of 0.1 MWh and maximum and minimum price of $\pm$\SI{9999}{EUR} on the EPEX Spot markets. The EPEX markets are anonymous and take place on every day of the year. A timeline of the daily routine of the German spot market is shown in Figure \ref{fig:daily_routine}.
In addition to the EPEX exchanges, it is also possible to trade with external partners, which is called over-the-counter (OTC) trading.
Since OTC trading is difficult to automate due to the lack of standardization of products, this work only considers trading on the EPEX exchange.

\footnotetext{On June 14, 2024 two more intraday auctions have been added. One at 22:00 on the day preceding the delivery and one at 10:00 on the delivery day for the remaining half of the day. This recent update is not considered for the work at hand.}

In order to optimally operate a BESS according to market signals, different strategies may be used. This work focuses primarily on the bidding and operation on the EPEX spot market in order to maximize the operator's revenues. 
The general framework for each strategy is that first input data is required, and then signals for trades are generated. 
The input data mainly includes price forecasts and asset data, such as the current battery state of charge (\acs{soc}). 
The price forecasts can either be done in-house or be obtained from a third-party provider; the asset data should be retrievable from internal databases. 
The generated signals can then be manually traded or transferred to an algorithm trader, which effectively executes the trades. 
An algorithm trader is preferable for IDC because the trading execution is not straightforward, considering that the trader's own offers also influence the market. 
For example, a buy position that is too large would drive the price up sharply, as the other market participants assume that the trader is in need of energy and therefore also willing to pay a higher price.
This is not equally the case for the DAA and IDA auctions, as they work according to the pay-as-cleared principle, and therefore the price shape is more relevant. In addition, the higher liquidity means that there is less price influence.
For the two auctions (DAA and IDA), this process would have to be carried out once a day shortly before bid submission. For the continuous market, however, this process should be repeated on an ongoing basis, at least every quarter-hour, to capture the entire market dynamics. 
With faster price forecasts or directly based on order book data, the optimization could even be performed far more frequently.

\begin{figure}
    \centering
    \includegraphics{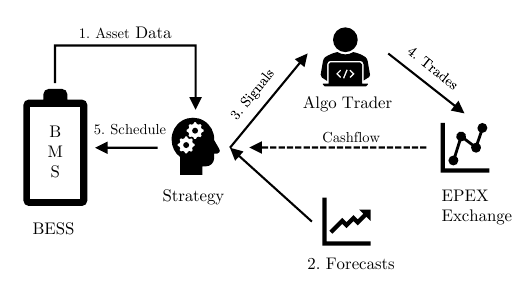}
    \caption{Illustration of a Generic BESS Operation Framework.}
    \label{fig:framework}
\end{figure}

The schedule resulting from the energy trading can then be automatically transferred to the Battery Management System (\acs{bms}) of the respective plant. The BMS includes both software and hardware to control the battery's operational conditions to prolong the lifetime, maintain safety, and state estimation for energy management purposes. The tasks of the BMS also include cell equalizing, \acs{soc} estimation, state of health estimation, and temperature management.
For more details on the technical operation of BESS the interested reader is referred to the work of Rouholamini et al. \cite{Rouhamini:2022}.
For the remainder of the work, a perfect BMS is assumed, meaning perfect schedule execution, SOC and state of health estimation. The described generic framework is illustrated in Figure \ref{fig:framework}.

%
%------------------------------------------------------------------------------------
%
\section{Backtest Engine for EPEX Spot Trading} \label{backtest_engine}

In order to evaluate a trading strategy based on the aforementioned framework, a backtest engine called BEICT was developed as part of the work. 
It is written object-oriented in Python, using mainly the packages Pandas \cite{pandas:2020} and Numpy \cite{harris:2020} for calculations and accounting, as well as Matplotlib \cite{Hunter:2007matplotlib} to display the results. The internally developed package Qpy was used to connect Iqony's SQL databases with the backtest engine, thus enabling the integrated import of the relevant market data.

To effectively transfer the described framework into a backtester, some assumptions had to be made for simplification. First of all, the Day Ahead Auction is not carried out in advance for each day but only based on the consecutive price results or price predictions of previous auctions for the length of the entire timespan. The same applies to the Intraday Auction. 
The ID1 index published by EPEX was used for the Intraday Continuous trading.
It is the volume-weighted average price of all continuous trades executed within the last trading hour of a contract.
The index was utilized due to several challenges associated with backtesting at the order book level. Firstly, the volume of data is substantial because of the order book's depth. Secondly, matching the trades with the corresponding offers adds another complexity. Thirdly, an algorithm capable of effectively executing these trades, similar to operations on the real exchange, would be required. In summary, the complexity of such an approach would exceed the project's scope.
Therefore, it was assumed that this index is tradable without liquidity constraints.
In addition, this approach for IDC trading neglects the trading period from (d-1) 16 to (d-1) 24 o'clock, which probably has a subordinate influence. 
Nevertheless, with a simulation duration of one day, there are almost no errors made.
Accordingly, it was also assumed that the bids were always accepted, with a tradable time period of 24 hours on IDC.
Transaction costs of the EPEX were also neglected.

The software basically consists of five objects called \texttt{DataLoader}, \texttt{TimeTravelHandler}, \texttt{EventManager}, \texttt{SignalGenerator}, \texttt{TradeSimulator} and \texttt{Portfolio}.
The \texttt{DataLoader} is responsible for importing and providing the relevant price data and forecasts from an SQL database via the Qpy package.
The \texttt{EventManager} manages different trading events such as the Day Ahead and Intraday Auctions, as well as the trading events for the intraday continuous market, which are done for every quarter for the predefined timeframe in a rolling horizon fashion.
The  \texttt{SignalGenerator} is used by the engine to create the signals according to the strategy that is tested. These signals are then converted and calculated into trades by the \texttt{TradeSimulator}.
The simulation of the backtesting period is controlled by the \texttt{TimeTravelHandler}. 
The resulting positions, trades, and cash flows are recorded in the \texttt{Portfolio} object.
New trading strategies must currently be hard-coded in the package as separate functions for the respective markets. 
As an alternative to the direct problem formulation, a generic framework, such as the EAO package \cite{pfingsten2021eao} can be employed in the backtester, especially when the strategy should involve a diverse portfolio of assets.

In the following, the program flow of BEICT is described:
Once the strategy to be tested has been implemented and the test period has been defined, the program can be started.
First, the required data is loaded from the database into the corresponding object. The program then begins with the DAA, whereby the hourly forecast for the entire test period is loaded first. The program then generates trading signals based on the forecast and the initial asset data using the predefined strategy. These signals are then offset against the real prices to simulate the trades. The resulting revenues and positions are then saved in the portfolio.
The next step is to proceed with the IDA, where the corresponding forecasts for the entire period are also retrieved first. The existing position from the DAA is then called up, which is forwarded together with the forecasts to create the IDA signals. These signals are also offset against the correct prices to simulate the trades. The resulting positions and revenues are stored in the portfolio again.

\begin{figure*}
    \centering
    \includegraphics[width=\textwidth]{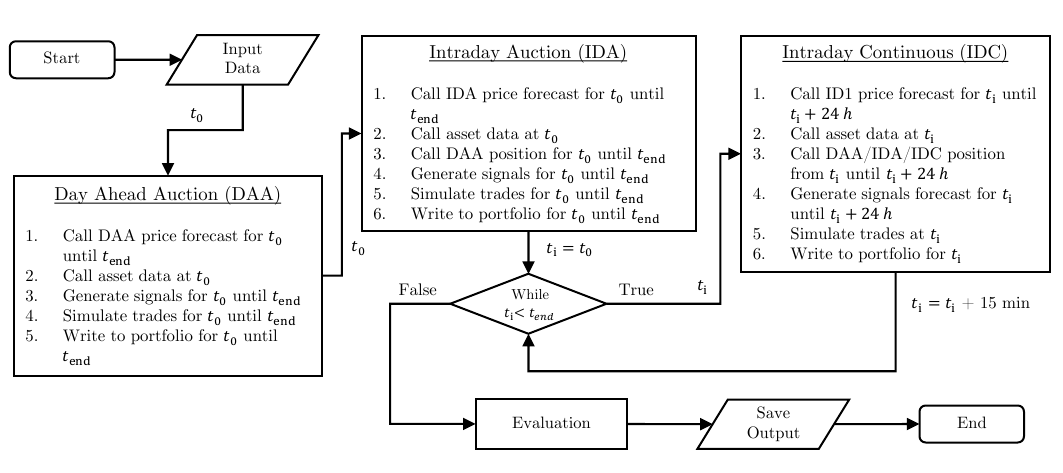}
    \caption{Flowchart of the Backtest Engine BEICT.}
    \label{fig:flowchart}
\end{figure*}

IDC trading is simulated slightly differently. In this case, IDC is called in a loop and thus executed in a rolling horizon fashion. 
For this purpose, an ID1 forecast for the next 24 hours is retrieved every quarter of an hour. 
In addition, the existing positions from DAA, IDA, and the already set IDC trades are retrieved. Also, the asset data for the current time is retrieved, and on this basis, IDC signals for the next 24 hours are generated. 
The corresponding trades are then recalculated against the actual ID1 index and saved in the portfolio. This process is repeated for every quarter of an hour until the end of the simulation. 
Afterwards, the results from the individual markets are evaluated.
Finally, the results of the backtesting are displayed or optionally saved as an Excel file and plotted.
A flowchart of the program sequence can be found in Figure \ref{fig:flowchart}.

%
%------------------------------------------------------------------------------------
%
\section{Mathematical Formulation} \label{stategy_formulation}

An example of an optimization-based strategy for battery marketing is described below. 
The proposed strategy builds on the work of CFP FlexPower GmbH (see \cite{flexpower:2024a} and \cite{flexpower:2024b}), but extends its model to include battery efficiency, self-discharge rate, a rolling horizon approach for IDC trading, and embedding it in the proposed operation framework.
In the following, three different optimizations are presented: one for the Day-Ahead Auction, one for the Intraday Auction, and the last for Intraday Continuous trading. 
The approach involves closing existing positions and resetting them if a subsequent market shift occurs.

\subsection{Day Ahead Auction (\acs{daa})}

The optimization problem for the Day Ahead Auction can be described by:
\begin{equation} \label{eq:objective_da}
\underset{\bm{x}}{\text{minimize}} \sum_{h=0}^{H} \widehat{c^{\daa}}[h] \cdot (P^{\daa}_{\charge}[h]-P^{\daa}_{\dis}[h]),
\end{equation}
\quad subject to
\begin{equation} \label{eq:E_max_da}
    0 \leq E[h] \leq E^{\max},
\end{equation}
\begin{equation} \label{eq:E_init_da}
    E[0] = E_{\init},
\end{equation}
\begin{multline} \label{eq:E_step_da}
    E[h+1] = E[h] \cdot (1-\gamma)^{\Delta t} \\ + (P^{\daa}_{\charge}[h] \cdot \eta^{\charge} - P^{\daa}_{\dis}[h] \cdot 1/\eta^{\dis}),
\end{multline}
\begin{equation} \label{eq:P_char_max_da}
    0 \leq P^{\daa}_{\charge}[h] \leq P^{\max} \cdot X,
\end{equation}
\begin{equation} \label{eq:P_dis_max_da}
    0 \leq P^{\daa}_{\dis}[h] \leq P^{\max} \cdot (1 - X),
\end{equation}
\begin{equation} \label{eq:volume_con_char_da}
    \Delta t \cdot \sum_{h=0}^{H} P^{\daa}_{\charge}[h] \cdot \eta^{\charge} \leq E^{\max} \cdot n^{\zyk} \cdot \frac{H}{24},
\end{equation}
\begin{equation} \label{eq:volume_con_dis_da}
    \Delta t \cdot \sum_{h=0}^{H} P^{\daa}_{\dis}[h]/\eta^{\dis} \leq E^{\max} \cdot n^{\zyk} \cdot \frac{H}{24}.
\end{equation}

The vector $\bm{x} = [\bm{P}^{\daa}_{\charge}, \bm{P}^{\daa}_{\dis}, \bm{E}, \bm{X}]^{\mathrm{T}}$ contains the decision variables, where  $\bm{P}\in\R^{H}$ and $\bm{E}\in\R^{H^{+}}$ are continuous variables that denote power and energy, respectively. The charge or discharge state of the BESS is indicated by the binary variable $\bm{X}\in\{0,1\}^{H}$.
The objective for this optimization is to maximize revenues from the DAA with utilization of DAA forecasts $\widehat{c^{\daa}}$, as displayed in the objective function in Equation \eqref{eq:objective_da}.
The following constraints describe the behavior of a BESS. First, the energy capacity is limited to $E^{\max}$ in Equation \eqref{eq:E_max_da} and an initial value is set in Equation \eqref{eq:E_init_da}. The state of charge change of the battery is described in \eqref{eq:E_step_da}, with the term including $\gamma$ to account for the battery's self-discharge rate. The charge and discharge power limits are described in \eqref{eq:P_char_max_da} and \eqref{eq:P_dis_max_da}, with the use of a binary variable $X$ to prevent simultaneous charging and discharging.
Without this binary variable, the storage system would charge and discharge simultaneously at negative prices in order to generate a profit from its efficiency losses, which is impossible in reality.
A full equivalent cycle-based charge and discharge volume limit can be set via constraints \eqref{eq:volume_con_char_da} and \eqref{eq:volume_con_dis_da}.
Equation \eqref{eq:E_max_da} is valid for all $h \in \{0,\ldots,H^{+}\}$, while \eqref{eq:E_step_da} to \eqref{eq:volume_con_dis_da} are valid for all $h \in \{0,\ldots,H\}$.
As the time resolution for this auction is hourly, the resulting charge and discharge powers can be directly transferred to the corresponding energy trades. 

\subsection{Intraday Auction (\acs{ida})}

To model the closing strategy, two variables for charging or buying ($P^{\ida}_{\charge},\overline{P^{\ida}_{\charge}}$) and discharging or selling ($P^{\ida}_{\dis},\overline{P^{\ida}_{\dis}}$) are introduced. While $P^{\ida}_{\charge}$ and $P^{\ida}_{\dis}$ are used to set new trades in the Intraday Auction, $\overline{P^{\ida}_{\charge}}$ and $\overline{P^{\ida}_{\charge}}$ are used to set trades to close existing positions from the DAA.
Based on this concept, the optimization problem for the Intraday Auction respecting the existing position from DAA with $P^{\daa}_{\charge}$ and $P^{\daa}_{\dis}$ (adjusted to quarter-hours) can be described by:
\begin{multline} \label{eq:objective_intra}
    \underset{\bm{x}}{\text{minimize}} \sum_{q=0}^{Q} \widehat{c^{\ida}}[q] \cdot \Delta t 
    \\ \cdot (P^{\ida}_{\charge}[q]-P^{\ida}_{\dis}[q]+ \overline{P^{\ida}_{\charge}}[q]- \overline{P^{\ida}_{\dis}}[q]),
\end{multline}
\quad subject to
\begin{equation} \label{eq:E_max_intra}
    0 \leq E[q] \leq E^{\max},
\end{equation}
\begin{equation} \label{eq:E_init_intra}
    E[0] = E_{\init},
\end{equation}
\begin{multline} \label{eq:E_step_intra}
    E[q+1] = E[q] \cdot (1-\gamma)^{\Delta t} \\
    +(P^{\ida}_{\charge}[q]+\overline{P^{\ida}_{\charge}}[q]+P^{\daa}_{\charge}[q]) \cdot \eta^{\charge} \cdot \Delta t \\
    - (P^{\ida}_{\dis}[q]+\overline{P^{\ida}_{\dis}}[q]+P^{\daa}_{\dis}[q]) \cdot \Delta t/\eta^{\dis},
\end{multline}
\begin{equation} \label{eq:P_char_max_intra}
    (P^{\ida}_{\charge}[q]+P^{\daa}_{\charge}[q]) \leq  P^{\max},
\end{equation}
\begin{equation} \label{eq:P_dis_max_intra}
    (P^{\ida}_{\dis}[q]+P^{\daa}_{\dis}[q]) \leq  P^{\max},
\end{equation}
\begin{multline} \label{eq:volume_con_char_intra}
    \Delta t \cdot \sum_{q=0}^{Q} (P^{\ida}_{\charge}[q]-\overline{P^{\ida}_{\dis}}[q] + P^{\daa}_{\charge}[q]) \cdot \eta^{\charge} \\ \leq E^{\max} \cdot n^{\zyk} \cdot \frac{Q}{96},
\end{multline}
\begin{multline} \label{eq:volume_con_dis_intra}
    \Delta t \cdot \sum_{q=0}^{Q} (P^{\ida}_{\dis}[q]-\overline{P^{\ida}_{\charge}}[q]+P^{\daa}_{\dis}[q])/\eta^{\dis} \\ \leq E^{\max} \cdot n^{\zyk} \cdot \frac{Q}{96},
\end{multline}
\begin{equation} \label{eq:da_dis_closing_logic}
    0 \leq \overline{P^{\ida}_{\charge}}[q] \leq P^{\daa}_{\dis}[q],
\end{equation}
\begin{equation} \label{eq:da_char_closing_logic}
    0 \leq \overline{P^{\ida}_{\dis}}[q] \leq P^{\daa}_{\charge}[q].
\end{equation}

Where \eqref{eq:objective_intra} describes the revenue-maximizing objective function and the BESS is described by Equations \eqref{eq:E_max_intra} to \eqref{eq:da_char_closing_logic}.
While constraints \eqref{eq:E_max_intra} and \eqref{eq:E_init_intra} are similar to the previous modeling, the energy step constraint \eqref{eq:E_step_intra} now includes both new variables and the existing positions from DAA as parameters.
The constraints \eqref{eq:P_dis_max_intra} and \eqref{eq:P_char_max_intra} account for the power limit of the storage, and \eqref{eq:volume_con_char_intra} and \eqref{eq:volume_con_dis_intra} limit the charge and discharge volume, respectively. In addition, the volume limits are responsible for the fact that the new positions can only be as high as the closed ones.
Finally, due to Equations \eqref{eq:da_dis_closing_logic} and \eqref{eq:da_char_closing_logic} the closing positions must correspond to the old positions.  
The vector $\bm{x} = [\bm{P}^{\ida}_{\charge}, \bm{P}^{\ida}_{\dis}, \overline{\bm{P}^{\ida}_{\charge}}, \overline{\bm{P}^{\ida}_{\dis}}, \bm{E}, \bm{X}]^{\mathrm{T}}$ contains the decision variables $\bm{P}\in\R^{Q}$ for power, $\bm{E}\in\R^{Q^{+}}$ for energy, and $\bm{X}\in\{0,1\}^{Q}$ for state indication.
Equation \eqref{eq:E_max_intra} is valid $\forall q \in \{0,\ldots,Q^{+}\}$, while Equations \eqref{eq:E_step_intra} to \eqref{eq:da_char_closing_logic} are valid $\forall q \in \{0,\ldots,Q\}$.
To extract the according trading signals from this strategy, the resulting charging and discharging powers must be multiplied by the time resolution $\Delta t$.

\subsection{Intraday Continuous (\acs{idc})}

The optimization problem for the Intraday continuous closing strategy is similar to the intraday auction, but it utilizes a rolling horizon framework. 
Meaning, the optimization is not performed for the entire timeframe, but for a window of the length $N_{p}$ for every time step of the simulation. 
In addition, for the purpose of simplification, the ID1 index for electricity prices is used. Using this simplification, the optimization can be described by:
\begin{multline} \label{eq:objective_contin}
    \underset{\bm{x}}{\text{minimize}} \sum_{q=q_{i}}^{q_{i}+N_{p}-1} \widehat{c^{\idc}}[q] \cdot \Delta t 
    \\\cdot (P^{\idc}_{\charge}[q]-P^{\idc}_{\dis}[q]+ \overline{P^{\idc}_{\charge}}[q]- \overline{P^{\idc}_{\dis}}[q]),
\end{multline}
\quad subject to
\begin{equation} \label{eq:e_max_contin}
    0 \leq E[q] \leq E^{\max},
\end{equation}
\begin{equation}
    E[q_{i}] = E^{\idc}_{\init},
\end{equation}
\begin{multline} \label{eq:volume_con_char_contin}
    E[q+1] = E[q] \cdot (1-\gamma)^{\Delta t}\\ 
    +(P^{\idc}_{\charge}[q]+\overline{P^{\idc}_{\charge}}[q]+P^{\daa,\ida,\idc}_{\charge}[q]) \cdot \eta^{\charge} \cdot \Delta t \\
    - (P^{\idc}_{\dis}[q]+\overline{P^{\idc}_{\dis}}[q]+P^{\daa,\ida,\idc}_{\dis}[q]) \cdot \Delta t/\eta^{\dis},
\end{multline}
\begin{equation}
    (P^{\idc}_{\charge}[q]+P^{\daa,\ida,\idc}_{\charge}[q]) \leq  P^{\max},
\end{equation}
\begin{equation}
    (P^{\idc}_{\dis}[q]+P^{\daa,\ida,\idc}_{\dis}[q]) \leq  P^{\max},
\end{equation}
\begin{multline}
    \sum_{q=q_{i}}^{q_{i}+N_{p}-1} (P^{\idc}_{\charge}[q]-\overline{P^{\idc}_{\dis}}[q] + P^{\daa,\ida,\idc}_{\charge}[q]) \\ \cdot \eta^{\charge} \cdot \Delta t
    \leq E^{\max} \cdot n^{\zyk} \cdot \frac{Q}{96},
\end{multline}
\begin{multline}
    \sum_{q=q_{i}}^{q_{i}+N_{p}-1} (P^{\idc}_{\dis}[q]-\overline{P^{\idc}_{\charge}}[q]+P^{\daa,\ida,\idc}_{\dis}[q]) \\ \cdot  1/\eta^{\dis} \cdot \Delta t
    \leq  E^{\max} \cdot n^{\zyk} \cdot \frac{Q}{96},
\end{multline}
\begin{equation}
    0 \leq \overline{P^{\idc}_{\charge}}[q] \leq P^{\daa,\ida,\idc}_{\dis}[q],
\end{equation}
\begin{equation} \label{eq:contin_char_closing_logic}
    0 \leq \overline{P^{\idc}_{\dis}}[q] \leq P^{\daa,\ida,\idc}_{\charge}[q].
\end{equation}

These equations are similar to the intraday auction ones, meaning Equation \eqref{eq:objective_contin} describes the objective function to maximize revenues and \eqref{eq:e_max_contin} to \eqref{eq:contin_char_closing_logic} describe the BESS behavior. In this rolling horizon framework, the initial SOC is the resulting SOC ($E^{\idc}_{\init}$) from the previous optimization. The existing position from the previous DAA and IDA, as well as previous continuous trades, is respected by $P^{\daa,\ida,\idc}_{\dis}$ and $P^{\daa,\ida,\idc}_{\charge}$. 
The vector $\bm{x} = [\bm{P}^{\idc}_{\charge}, \bm{P}^{\idc}_{\dis}, \overline{\bm{P}^{\idc}_{\charge}}, \overline{\bm{P}^{\idc}_{\dis}}, \bm{E}, \bm{X}]^{\mathrm{T}}$ contains the decision variables $\bm{P}\in\R^{N_{p}}$ for power,$\bm{E}\in\R^{N_{p}+1}$ for energy, and $\bm{X}\in\{0,1\}^{N_{p}}$ again for state indication.
Equation \eqref{eq:e_max_contin} is valid $\forall q \in \{q_{i},\ldots,q_{i}+N_{p}\}$, while equations \eqref{eq:volume_con_char_contin} to \eqref{eq:contin_char_closing_logic} are valid $\forall q \in \{q_{i},\ldots,q_{i}+N_{p}-1\}$.
To extract the according trading signals from this strategy, the resulting charging and discharging powers must be multiplied by the time resolution.

The strategy described above was implemented using Python with the modelling language Pyomo \cite{pyomo:2011} in combination with the solver Gurobi \cite{gurobi} and GLPK \cite{Oki2012GLPKL} in the developed backtest engine BEICT.
Its functionality is tested in the following case study.

%
%------------------------------------------------------------------------------------
%
\section{Case Study} \label{case_study}

In the subsequent case study, an analysis is conducted to determine the potential earnings of a BESS over the past few years. 
A \SI{10}{\MW} and \SI{10}{\MWh} standalone grid-connected BESS in Germany was assumed that wants to participate in energy arbitrage.
A battery based on LFP cells with a charging and discharging efficiency of 95 \% was adopted. The self-discharge rate is set at 3 \% per month, and a maximum of two full equivalent cycles per day are allowed. The prediction horizon for IDC trading was set to 24 hours and the initial SOC to zero.
The revenue generated by the system operating with the aforementioned strategy over the past five years is examined.
In this first study, a perfect forecast is assumed, i.e. the forecast corresponds to the actual prices. The analysis therefore initially shows an upper limit for revenues.
In addition to the already mentioned simplifications of the backtest engine, all taxes and levies are neglected.

\subsection{Results and Discussion}

\begin{figure}[htp]
    \centering
    \includegraphics{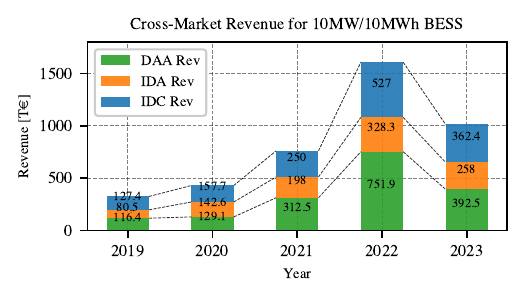}
    \caption{Cross-Market Revenue of a 10MW/10MWh BESS from 2019 until 2023.}
    \label{fig:revenues}
\end{figure}

Table \ref{table:auswertung} presents the annual revenue generated by a \SI{10}{\MW}/\SI{10}{\MWh} standalone BESS in Germany, engaged in energy arbitrage across three market segments: Day Ahead Auction, Intraday Auction, and Intraday Continuous. 
The results span from 2019 to 2023, showing both the total revenue and the revenue share from each market are visualized in Figure \ref{fig:revenues}.
In 2023, the BESS generated a total revenue of \num{1013.0} TEUR, with DAA contributing 392.5 TEUR (38.7\%), IDA contributing 258.0 TEUR (25.5\%), and IDC contributing 362.4 TEUR (35.8\%). 
The highest total revenue was observed in 2022 at \num{1607.2} TEUR, predominantly driven by a significant contribution from the DAA market of 751.9 TEUR (46.8\%). 
Conversely, 2019 recorded the lowest total revenue at 324.4 TEUR, with IDC constituting the largest share with 127.4 TEUR (39.3\%).
Over the five-year period, the average annual revenue was around 827 TEUR, with DAA, IDA, and IDC contributing 38.5\%, 26.0\%, and 35.5\%, respectively. 

These figures point to the dynamic nature of energy arbitrage markets, where revenue distribution among DAA, IDA, and IDC varies significantly each year due to fluctuating market conditions.
The observed revenue volatility is primarily driven by the underlying price volatility in the energy markets. 
As electricity price volatility has increased lately, the potential for revenue from arbitrage operations has similarly risen. 
This trend is reflected in the significant jump in revenues, particularly notable in the high earnings of 2022 that were caused by extreme price volatility resulting from the Russian attack on Ukraine. 

\begin{table}[htp]
    \caption{Cross-Market BESS Revenue from 2019 until 2023.}
    \centering
    %\small
    \begin{tabular}{l|ccccc|c} 
    \toprule
    Year & 2023 & 2022 & 2021 & 2020 & 2019 & Avg. \\ \midrule
     DAA Rev. [T€] & \num{393} & \num{752} & \num{313} & \num{129} & \num{116} & \num{341}\\
     - Share [\%] & 38.7 & 46.8 & 41.1 & 30,1 & 35.9 & 38.5\\
     IDA Rev. [T€] & \num{258} & \num{328} & \num{198} & \num{143} & \num{81} & \num{202}\\
     - Share [\%] & 25.5 & 20.4 & 26.0 & 33.2 & 24.8 & 26.0\\ 
     IDC Rev. [T€] & \num{362} & \num{527} & \num{250} & \num{158} & \num{127} & \num{285}\\
     - Share [\%] & 35.8 & 32.8 & 32.9 & 36.7 & 39.3 & 35.5\\ \midrule
     Total Rev. [T€] & \num{1013} & \num{1607} & \num{760} & \num{430} & \num{324} & \num{827}\\ \bottomrule
    \end{tabular}
    \label{table:auswertung}
\end{table}

\begin{figure}
    \centering
    \includegraphics{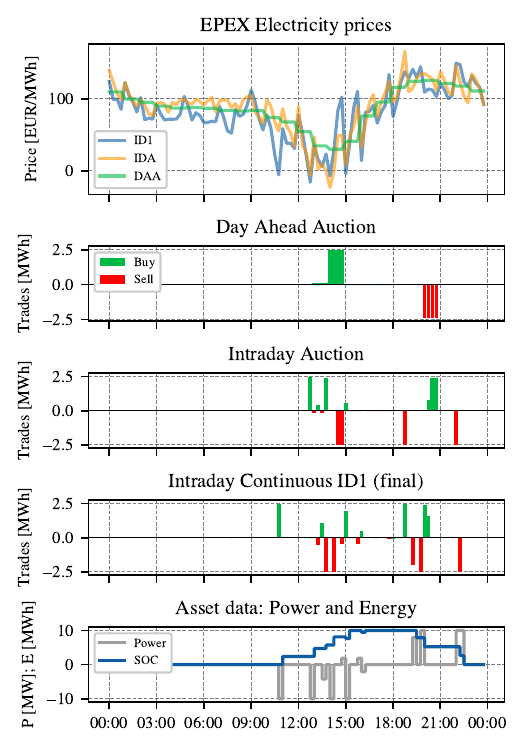}
    \caption{BEICT Example Backtesting Results for 01/05/2023.}
    \label{fig:Results}
\end{figure}

Figure \ref{fig:Results} illustrates the backtesting results for the standalone BESS operating on May 1, 2023. 
This figure provides an intricate hourly breakdown of electricity prices, trades executed in different market segments (DAA, IDA, and IDC), and BESS power and SOC throughout the day.

The top panel of Figure \ref{fig:Results} reveals the (quarter-) hourly electricity prices for the DAA, IDA, and IDC markets. 
The prices exhibit marked volatility, characteristic of the dynamic nature of the electricity market as an interplay of demand and supply. 
This volatility enables the BESS to exploit price differentials across time and different market segments.
The second panel delineates the BESS's buy and sell trades executed in the DAA market. 
Represented as vertical bars, these trades indicate the volume of energy transactions per hour.
It can be seen that the two main trades happen at the maximum and minimum DAA price.
The third panel presents the trades in the IDA market, depicted as vertical bars for each quarter-hour interval. 
The IDA trades serve to adjust positions taken in the DAA market, responding to updated price forecasts and evolving market conditions throughout the trading day.
In addition, further available energy can be marketed here.
The fourth panel illustrates the trades in the IDC market, showing frequent buying and selling activities at quarter-hour intervals. The continuous nature of this market allows the BESS to react swiftly to real-time price fluctuations, continuously optimizing its trading positions to enhance revenue. 
The agility demonstrated in this market segment underscores the importance of real-time market monitoring and rapid decision-making capabilities. As the trades done in the IDC are updated every quarter, in this chart only the final net trades are shown.
The bottom panel tracks the SOC of the BESS over the 24-hour period, displaying both the SOC and the corresponding power levels. The SOC fluctuates as the BESS alternates between charging and discharging in response to market opportunities. In final results, the battery charges during afternoon low prices and discharges at high evening prices.

\subsection{Sensitivity Analysis}

\begin{figure}[htp]
    \centering
    \includegraphics[width=.5\textwidth]{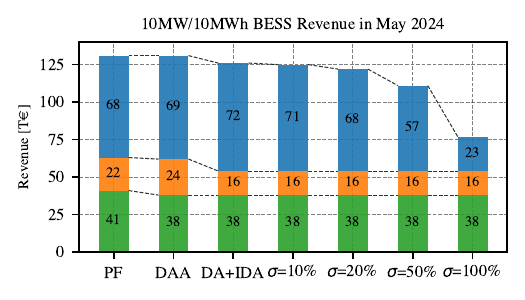}
    \caption{Cross-Market BESS Revenue Influence of Forecast Uncertainty in May 2024.}
    \label{fig:sensitivity}
\end{figure}

The subsequent sensitivity analysis examines the extent to which the inclusion of real forecasts or prediction errors affects the performance in comparison to a 100 \% accurate forecast.
For this purpose, the DAA forecast was initially replaced by a real daily forecast from ICIS \cite{ICIS:2024}, followed by the IDA forecast. 
The provider combines various models into an ensemble model for the electricity price forecast, which consists of a MILP, LP, \acs{nn} or uses fundamental models of the power market.
As no ID1 forecast was available, a volatility of 10, 20, 50 and 100 \% was manually applied to the ID1 index to simulate forecast errors.
A clean data series was only available from 04/05/2024 to 04/06/2024, so this period was used for backtesting. 
The Mean Absolute Error (\acs{mae}) for the ICIS DAA forecast in this period was \num{10.65} and the Root Mean Square Error (\acs{rmse}) \num{15.48}. 
A MAE value of \num{14.94} and a RMSE of \num{20.91} were determined for the ICIS IDA forecast. 
The backtesting results using the described forecasts can be found in the following Table \ref{table:sensitivity} and are visualized in Figure \ref{fig:sensitivity}.
\begin{table}
    \caption{Cross-Market BESS Revenue with Forecast Uncertainty.}
    \centering
    %\small
    \begin{tabular}{l|ccc|c} 
    \toprule
    Revenue from [T€] & DAA & IDA & IDC & Total \\ \midrule
    \textit{Perfect Forecast (PF)} & 41.4 & 22.4 & 67.7 & 131.5\\
    ICIS DAA & 38.4 & 24.5 & 69.2 & 132.1\\
    ICIS DAA + IDA & 38.4 & 16.1 & 72.0 & 126.5\\ \midrule
    ID1 ($\sigma=10\%$) & 38.4 & 16.1 & 71.2 & 125.7\\
    ID1 ($\sigma=20\%$) & 38.4 & 16.1 & 68.0 & 122.5\\
    ID1 ($\sigma=50\%$) & 38.4 & 16.1 & 57.3 & 111.8\\
    ID1 ($\sigma=100\%$) & 38.4 & 16.1 & 22.6 & 77.1\\
    \bottomrule
    \end{tabular}
    \label{table:sensitivity}
\end{table}
Using perfect forecasts, the BESS generated a total revenue of 131.5 TEUR in the test period, with DAA contributing 41.4 TEUR (31\%), IDA contributing 22.4 TEUR (17\%) and IDC contributing 67.7 TEUR (52\%). Compared to the yearly averages before, the amount generated by the IDC is comparatively high. When utilizing the ICIS DAA forecast, the DAA revenues fall about 7\% to 38.4 TEUR. However, in this case both on the IDA and IDC a 9\% and 2\% higher revenue could be achieved, resulting in a higher total revenue of 132.1 TEUR.
When using the ICIS IDA forecast in addition, the respective revenue drops about 34\% from 24.5 TEUR to 16.1 TEUR. 
The IDC revenue increases slightly, but the total revenue falls to 126.5 TEUR.
With the added volatility of 10\% the IDC revenue falls only about 0.8 TEUR or 1\%. A 20\% increase in volatility caused a 6\% drop from 72 TEUR to 68 TEUR. Even the 50\% added volatility only resulted in a revenue decrease to 57.3 TEUR or 20\%. Finally with 100\% increased volatility, the IDC revenue decreased to 22.6 TEUR about 69\% from the perfect foresight result.
The findings indicate that a substantial portion of the outcomes can be derived from authentic forecasts.

%
%------------------------------------------------------------------------------------
%
\section{Conclusion and Outlook} \label{Conclusion}

In this work, the optimal operation of standalone BESS in the cross-market energy arbitrage business was addressed. For this purpose, a generic framework for energy trading integrated BESS operation was presented first. Afterwards, the developed backtest engine for this work was described. Then an enhanced operating strategy formulation was presented and subsequently tested in a case study with historical prices of the last five years. In addition, the influence of forecasting error was investigated in a sensitivity analysis.
The novel strategy incorporating battery efficiency, self-discharge, and a rolling horizon approach for IDC trading proved successful in the conducted case study. 
The case study has shown significantly rising BESS revenues in recent years, with a clear outlier in 2022 due to extreme electricity price volatility following the Russian attack on Ukraine.
If this trend persists, it is feasible that BESS will demonstrate favorable economic efficiency in arbitrage.
The subsequent sensitivity analysis showed that the real prediction of the DAA did not significantly influence the result. In contrast, the real prediction for the IDA had a stronger influence. The IDC at least proved to be resistant to increased volatility, so that stronger declines were only recorded from a 50\% increase in volatility.

Both the backtester developed, as well as the strategy and the case study carried out have limitations mainly resulting from the assumptions mentioned. 
For example, important market aspects such as liquidity and bid-offer spread were neglected.
Future work could improve on these limitations. For example, the DAA and IDA trades could be simulated daily in rolling horizon fashion in the backtester, which will also result in a more realistic trading strategy.
In addition, for the strategy, the two new intraday auctions could be included, which probably will increase revenue due to more arbitrage opportunities. 
The case study could be improved by also investigating the influence of a real forecast for the ID1 index. 
The concept could also be validated with a real algorithm trader to investigate the strategy on the order book level and the influence of market liquidity.

\ifCLASSOPTIONcaptionsoff
  \newpage
\fi

\bibliographystyle{IEEEtran}
\bibliography{bibtex/bib/Literatur}

\begin{thebibliography}{10}
\providecommand{\url}[1]{#1}
\csname url@samestyle\endcsname
\providecommand{\newblock}{\relax}
\providecommand{\bibinfo}[2]{#2}
\providecommand{\BIBentrySTDinterwordspacing}{\spaceskip=0pt\relax}
\providecommand{\BIBentryALTinterwordstretchfactor}{4}
\providecommand{\BIBentryALTinterwordspacing}{\spaceskip=\fontdimen2\font plus
\BIBentryALTinterwordstretchfactor\fontdimen3\font minus \fontdimen4\font\relax}
\providecommand{\BIBforeignlanguage}[2]{{%
\expandafter\ifx\csname l@#1\endcsname\relax
\typeout{** WARNING: IEEEtran.bst: No hyphenation pattern has been}%
\typeout{** loaded for the language `#1'. Using the pattern for}%
\typeout{** the default language instead.}%
\else
\language=\csname l@#1\endcsname
\fi
#2}}
\providecommand{\BIBdecl}{\relax}
\BIBdecl

\bibitem{Kurzweil:2018}
P.~Kurzweil and O.~K. Dietlmeier, \emph{Elektrochemische Speicher: Superkondensatoren, Batterien, Elektrolyse-Wasserstoff, Rechtliche Rahmenbedingungen}, 2nd~ed.\hskip 1em plus 0.5em minus 0.4em\relax Springer Medien Wießbaden.

\bibitem{Iqony:2023a}
\BIBentryALTinterwordspacing
{Speicher: Für eine funktionierende Energiewende}. Iqony GmbH. Accessed: Aug. 9, 2023. [Online]. Available: \url{https://www.iqony.energy/wachstumsprojekte/speicher}
\BIBentrySTDinterwordspacing

\bibitem{Iqony:2024a}
\BIBentryALTinterwordspacing
{Speicher: Flexibilität für eine erfolgreiche Energiewende}. Iqony GmbH. Accessed: Jul. 7, 2024. [Online]. Available: \url{https://www.iqony.energy/standortentwicklung/projekt-steady-green-energy}
\BIBentrySTDinterwordspacing

\bibitem{Lehmann:2022}
D.~Lehmann, D.~H. Rodriguez, and M.~Brack, ``Optimized operation of large scale battery systems,'' \emph{at - Automatisierungstechnik}, vol.~70, no.~1, pp. 67--78, 2022.

\bibitem{Lehmann:2018}
D.~Lehmann, M.~Mühl, P.~Deeskow, and H.~Yilmaz, ``Optimized {O}peration of {L}arge {S}cale {B}attery {S}ystems - {E}xperiences, {N}ew {O}pportunities and {B}ig {D}ata.''\hskip 1em plus 0.5em minus 0.4em\relax Wien, Austria: Electrify Europe Conference, 2018.

\bibitem{Hidalgo:2020}
D.~H. Rodríguez, D.~Lehmann, M.~Schmoltizi, and M.~Brack, ``Betriebserfahrung und {O}ptimierung von {G}roßbatteriesystemen: {E}ine {S}imulationstudie zur {O}ptimierung des {SoC}-managements.''\hskip 1em plus 0.5em minus 0.4em\relax KELI - Konferenz zur Elektro-, Leit- und Informationstechnik, 2020.

\bibitem{vanSandbergen:2024b}
L.~van Sandbergen, D.~Hidalgo~Rodriguez, and M.~Roemmich, ``{Optimized Sizing of Battery Energy Storage Systems in the Energy Arbitrage Business for Industrial Customers},'' in \emph{PESS 2024 – IEEE Power and Energy Student Summit}.\hskip 1em plus 0.5em minus 0.4em\relax VDE, 2024, pp. 85--90.

\bibitem{Collath:2023}
N.~Collath, M.~Cornejo, V.~Engwerth, H.~Hesse, and A.~Jossen, ``Increasing the lifetime profitability of battery energy storage systems through aging aware operation,'' \emph{Applied Energy}, vol. 348, p. 121531, 2023.

\bibitem{Kumtepeli:2020}
V.~Kumtepeli, H.~C. Hesse, M.~Schimpe, A.~Tripathi, Y.~Wang, and A.~Jossen, ``Energy {A}rbitrage {O}ptimization {W}ith {B}attery {S}torage: {3D-MILP} for {E}lectro-{T}hermal {P}erformance and {S}emi-{E}mpirical {A}ging {M}odels,'' \emph{IEEE Access}, vol.~8, pp. 204\,325--204\,341, 2020.

\bibitem{Metz:2018}
D.~Metz and J.~T. Saraiva, ``Use of battery storage systems for price arbitrage operations in the 15- and 60-min {G}erman intraday markets,'' \emph{Electric Power Systems Research}, vol. 160, pp. 27--36, 2018.

\bibitem{Hashmi:2019}
M.~U. Hashmi, A.~Mukhopadhyay, A.~Bušić, J.~Elias, and D.~Kiedanski, ``Optimal {S}torage {A}rbitrage under {N}et {M}etering using {L}inear {P}rogramming,'' in \emph{2019 IEEE International Conference on Communications, Control, and Computing Technologies for Smart Grids (SmartGridComm)}, 2019, pp. 1--7.

\bibitem{Xie:2021}
Y.~Xie, W.~Guo, Q.~Wu, and K.~Wang, ``Robust {MPC}-based bidding strategy for wind storage systems in real-time energy and regulation markets,'' \emph{International Journal of Electrical Power \& Energy Systems}, vol. 124, p. 106361, 2021.

\bibitem{Krishnamurthy:2018}
D.~Krishnamurthy, C.~Uckun, Z.~Zhou, P.~R. Thimmapuram, and A.~Botterud, ``Energy {S}torage {A}rbitrage {U}nder {D}ay-{A}head and {R}eal-{T}ime {P}rice {U}ncertainty,'' \emph{IEEE Transactions on Power Systems}, vol.~33, no.~1, pp. 84--93, 2018.

\bibitem{Abdulla:2018}
K.~Abdulla, J.~de~Hoog, V.~Muenzel, F.~Suits, K.~Steer, A.~Wirth, and S.~Halgamuge, ``Optimal {O}peration of {E}nergy {S}torage {S}ystems {C}onsidering forecasts and {B}attery {D}egradation,'' \emph{IEEE Transactions on Smart Grid}, vol.~9, no.~3, pp. 2086--2096, 2018.

\bibitem{Hafiz:2020}
F.~Hafiz, M.~A. Awal, A.~R.~d. Queiroz, and I.~Husain, ``Real-{T}ime {S}tochastic {O}ptimization of {E}nergy {S}torage {M}anagement {U}sing {D}eep {L}earning-{B}ased {F}orecasts for {R}esidential {PV} {A}pplications,'' \emph{IEEE Transactions on Industry Applications}, vol.~56, no.~3, pp. 2216--2226, 2020.

\bibitem{Pelzer:2016}
D.~Pelzer, D.~Ciechanowicz, and A.~Knoll, ``Energy arbitrage through smart scheduling of battery energy storage considering battery degradation and electricity price forecasts,'' in \emph{2016 IEEE Innovative Smart Grid Technologies - Asia (ISGT-Asia)}, 2016, pp. 472--477.

\bibitem{Abramova:2021}
E.~Abramova and D.~Bunn, ``Optimal {D}aily {T}rading of {B}attery {O}perations {U}sing {A}rbitrage {S}preads,'' \emph{Energies}, vol.~14, no.~16, 2021.

\bibitem{Loew:2021}
S.~Loew, A.~Anand, and A.~Szabo, ``Economic model predictive control of {L}i-ion battery cyclic aging via online rainflow-analysis,'' \emph{Energy Storage}, vol.~3, no.~3, p. e228, 2021.

\bibitem{NARAJEWSKI:2022}
M.~Narajewski and F.~Ziel, ``Optimal bidding in hourly and quarter-hourly electricity price auctions: Trading large volumes of power with market impact and transaction costs,'' \emph{Energy Economics}, vol. 110, p. 105974, 2022.

\bibitem{NEMO:2024}
\BIBentryALTinterwordspacing
{EUPHEMIA Public Description: Single Price Coupling Algorithm}. NEMO Comittee. Accessed: Aug. 17, 2024. [Online]. Available: \url{https://www.nemo-committee.eu/assets/files/euphemia-public-description.pdf}
\BIBentrySTDinterwordspacing

\bibitem{Baule:2021}
\BIBentryALTinterwordspacing
R.~Baule and M.~Naumann, ``{Volatility and Dispersion of Hourly Electricity Contracts on the German Continuous Intraday Market},'' \emph{Energies}, vol.~14, no.~22, 2021. [Online]. Available: \url{https://www.mdpi.com/1996-1073/14/22/7531}
\BIBentrySTDinterwordspacing

\bibitem{Rouhamini:2022}
M.~Rouholamini, C.~Wang, H.~Nehrir, X.~Hu, Z.~Hu, H.~Aki, B.~Zhao, Z.~Miao, and K.~Strunz, ``A {R}eview of {M}odeling, {M}anagement, and {A}pplications of {G}rid-{C}onnected {L}i-{I}on {B}attery {S}torage {S}ystems,'' \emph{IEEE Transactions on Smart Grid}, vol.~13, no.~6, pp. 4505--4524, 2022.

\bibitem{pandas:2020}
\BIBentryALTinterwordspacing
{The pandas development team}, ``pandas-dev/pandas: Pandas,'' Feb. 2020. [Online]. Available: \url{https://doi.org/10.5281/zenodo.3509134}
\BIBentrySTDinterwordspacing

\bibitem{harris:2020}
C.~R. Harris, K.~J. Millman, S.~J. van~der Walt, R.~Gommers, P.~Virtanen, D.~Cournapeau, E.~Wieser, J.~Taylor, S.~Berg, N.~J. Smith, R.~Kern, M.~Picus, S.~Hoyer, M.~H. van Kerkwijk, M.~Brett, A.~Haldane, J.~F. del R{\'{i}}o, M.~Wiebe, P.~Peterson, P.~G{\'{e}}rard-Marchant, K.~Sheppard, T.~Reddy, W.~Weckesser, H.~Abbasi, C.~Gohlke, and T.~E. Oliphant, ``Array programming with {NumPy},'' \emph{Nature}, vol. 585, no. 7825, pp. 357--362, Sep. 2020.

\bibitem{Hunter:2007matplotlib}
J.~D. Hunter, ``{Matplotlib: A 2D graphics environment},'' \emph{Computing in Science \& Engineering}, vol.~9, no.~3, pp. 90--95, 2007.

\bibitem{pfingsten2021eao}
\BIBentryALTinterwordspacing
T.~Pfingsten and D.~Oeltz, ``{EAO — A Framework for Optimizing Decentralized Portfolios and Green Supply},'' \emph{Fraunhofer Institute for Algorithms and Scientific Computing SCAI}, 2021. [Online]. Available: \url{https://ssrn.com/abstract=3842822}
\BIBentrySTDinterwordspacing

\bibitem{flexpower:2024a}
\BIBentryALTinterwordspacing
{FlexIndex}. CFP FlexPower GmbH. Accessed: Jul. 7, 2024. [Online]. Available: \url{https://flex-power.energy/services/flex-trading/flex-index/}
\BIBentrySTDinterwordspacing

\bibitem{flexpower:2024b}
\BIBentryALTinterwordspacing
{bess-optimizer}. CFP FlexPower GmbH. Accessed: Jul. 7, 2024. [Online]. Available: \url{https://github.com/FlexPwr/bess-optimizer}
\BIBentrySTDinterwordspacing

\bibitem{pyomo:2011}
W.~E. Hart, J.-P. Watson, and D.~L. Woodruff, ``Pyomo: modeling and solving mathematical programs in python,'' \emph{Mathematical Programming Computation}, vol.~3, no.~3, pp. 219--260, 2011.

\bibitem{gurobi}
\BIBentryALTinterwordspacing
{Gurobi Optimization, LLC}, ``{Gurobi Optimizer Reference Manual},'' 2023. [Online]. Available: \url{https://www.gurobi.com}
\BIBentrySTDinterwordspacing

\bibitem{Oki2012GLPKL}
\BIBentryALTinterwordspacing
A.~Makhorin, ``{GLPK (GNU Linear Programming Kit)},'' 2012. [Online]. Available: \url{https://www.gnu.org/software/glpk/}
\BIBentrySTDinterwordspacing

\bibitem{ICIS:2024}
\BIBentryALTinterwordspacing
{EU Power short-term forecast}. Independent Commodity Intelligence Services (ICIS). Accessed: Aug. 6, 2024. [Online]. Available: \url{https://www.icis.com/explore/}
\BIBentrySTDinterwordspacing

\end{thebibliography}

\end{document}